%Paper: hep-th/9301089
%From: drg@theory3.caltech.edu (Doron Gepner)
%Date: Thu, 21 Jan 93 19:27:28 PST

%%%%%%%%%%%%%%%%%%%%%%%%%%%%%%%%%%%%%%%%%%%%%%%%%%%%%%%%%%%
%                                                         %
%%%%%%%%%%%%%%%%%%%%%%%%%%%%%%%%%%%%%%%%%%%%%%%%%%%%%%%%%%%
\input phyzzx
\input mydef
\def\half{{1\over2}}

 %% should be defined.

\overfullrule=10pt
\date{December 1987}
\titlepage
\title{String Theory on Calabi--Yau Manifolds:}
\vskip20pt
\titlestyle{The Three Generations Case}
\author{Doron Gepner\footnote{\dagger}{Research
supported in part by NSF grant PHY-80-19754.}}
\JHL
\vskip15pt
\abstract
Recently, string theory on Calabi--Yau manifolds was constructed
and was shown to be a fully consistent, space--time supersymmetric
string theory. The physically interesting case is the case of three
generations. Intriguingly, it appears at the present that there is a
unique manifold which gives rise to three generations. We describe in this
paper a full fledged string theory on this manifold in which
the complete spectrum and all the
Yukawa couplings can be computed exactly.
\endpage
String theory is a rarity among physical theories. For twenty years
it has been developed without experimental input. Certainly,
it is a beautifully consistent theory. Yet, it remained the theoretician
playground.
\par
The purpose of this note is to make a step towards the confrontation
of string theory with experimental physics as a candidate theory for
the unification of all natural forces. We do this by exploring the
physically relevant case, the case of three generations, in the
vast framework of the recently
discovered
\REF\one{D. Gepner,
Princeton Univ. preprint, PUPT-1056, April, 1987,
and Nucl. Phys. B (in press)}\REF\two{D. Gepner, Princeton U. preprint
PUPT-1066, August (1987), and Phys. Lett. B (in press)}\r{\one,\two}\
string theory on Calabi--Yau manifolds.
\par
String theories incorporate an elaborate and tightly woven
structure with many consistency requirements. In particular,
a superstring propagating in flat Minkowski space is a consistent theory only
if this space is ten dimensional. The real world is however four
dimensional.
\par
A possible solution to this problem is
to consider a world which is a
manifold $M\times K$, where $M$ is the usual four dimensional
Minkowski space and $K$ is some tiny, `invisible', manifold,
an idea first suggested in the late twenties by Kaluza and Klein
\REF\KK{Th. Kaluza, Sitz.Preuss.Akad.Wiss.
K1 (1921), 966;  O. Klein, Z.Phys. 37 (1926), 895}\r\KK.
The
internal manifold $K$ would then give rise to the observable forces
and the particular choice for the manifold has a profound influence
on the physical predictions of the theory.
\par
To implement the Kaluza--Klein idea in string theory, one needs
to study string propagation on the manifold $M\times K$, where $K$ is
some curved internal manifold.
The first study of string propagation
in curved space was described in \REF\Group{D. Gepner and E. Witten,
Nucl. Phys. B278, 493 (1986)}\r\Group , where a closed bosonic string
theory
on a manifold $K$, which is a Lie group manifold, was considered. As
a result,
string theory on a group manifold was shown to be a fully consistent,
full fledged string theory, obeying all the severe constrains that
string theory should obey. The constraints of the existence of sensible
vertex operators, which are in one to one correspondence with the physical
spectrum, unitarity at the tree level
and at the one loop level, were all shown to be obeyed. Moreover, as a result
of this work, it became clear that a full fledged string theory can be
constructed, along the same lines, from any conformal field theory
in two dimensions.
\par
For phenomenological reasons, one would actually like to study
the case where the string theory has space--time supersymmetry. This
symmetry between fermions and bosons enables the elimination of
tachyons
and the potential resolution of questions like the vanishing
of the cosmological
constant and the hierarchy problem.
\par
The problem of the existence of supersymmetric string theory in
curved space was open for a long time. Initial interest in the
question came from the work of \REF\Candelas{P. Candelas, G. Horowicz,
A. Strominger and E. Witten, Nucl. Phys. B258, 46 (1985)}\r\Candelas ,
in which the equations of classical ten dimensional supergravity were studied
and were shown to have a solution, provided the manifold $K$
is a complex manifold
of vanishing first Chern class (Calabi--Yau manifold). This gives
initial indication that string theory on such a manifold might
exist, since the same classical equations
expresses the
lowest order contribution to the conformal anomaly of the sigma
model on the manifold \r\Candelas, which describes string propagation on it.
However, the tiniest conformal anomaly renders such a string theory
inconsistent so one needs to study higher order contributions.
Indeed, such a contribution in the four loop level was reported
in \REF\Grisaru{M. T. Grisaru, A. Van de Ven, and D. Zanon, Phys. Lett
173B, 423 (1986); D. Gross and E. Witten, Nucl. Phys. B277
(1986), 1}\r\Grisaru , implying that the naive string theory
on a Calabi Yau (CY) manifold is inconsistent. Arguments were suggested
\REF\Partial{D. Nemeshansky and A. Sen, Phys. Lett. 178B (1986), 365; M. Dine,
N. Seiberg, X.G. Wen and E. Witten, Nucl.Phys. 278B (1986), 769, 289B (1987)
319}\r\Partial\ that it may be possible to modify the metric and
many workers in the field believed that, at least for large radius,
a nearby string theory exists.
\par
This is not the only problem.
The existence of metrics of $SU(3)$ holonomy on Calabi Yau
manifolds was conjectured by Calabi and proved by Yau
\REF\YAU{S.T. Yau, Proc.Natl.Acad.Sci. 74 (1977), 1798}\r\YAU.
However, no metric of a compact
Calabi Yau manifold is known explicitly. The writing of such a
metric is a hard mathematical problem, which even for the simplest
$K3$ surface has been open for decades. Consequently, the propagation
of a classical Newtonian particle on such a manifold is untractable, since
one cannot compute distances.  Thus, it appears to be entirely hopeless
to compute the
physical predictions from the seemingly much harder problem of the
propagation of a quantized string on such a
manifold.
\par
In a recent series of papers \r{\one,\two}\ the author has
put forward
the construction of string theory on Calabi Yau manifolds and have shown
them to be fully consistent, space--time supersymmetric string theories,
where all the physical predictions can be computed exactly.
\par
The new idea, which avoids all the aforementioned difficulties, is
to proceed in two stages. At the first stage any possible
geometrical interpretation was ignored and new
space--time supersymmetric string theories were constructed
from scratch \r\one, solving
the stringent constraints that string theory must obey.
At the second stage, by studying the massless spectrum
of these new string theories and comparing it with the results
of equivariant index theorems in particular Calabi Yau
geometries, it was shown \r\two\ that these theories
correspond to string propagation on Calabi--Yau manifolds.
The logic behind this procedure is
that since string theory is a rarely constrained
system, it should be possible to recover any such theory
by simply considering string
theory and its constraints, per se.
This proves the consistency and
existence of CY string theories, as well as giving their actual
construction.
\par
To carry out the first stage it was needed to understand
how to construct space--time supersymmetric string theories.
Previously, the only known method to get space--time supersymmetry
in string theory
was in the context of theories made entirely out of free fermions,
along the lines of the GSO construction \REF\GSO{F. Gliozzi, D. Olive,
and J. Sherk, Nucl. Phys. B122 (1977), 253}
\r\GSO . In order to get supersymmetry one needs some
projection.
The problem is that
modular invariance
almost always prevents one from projecting out any fields
in a general conformal field theory. It is a kind
of completeness condition. Thus it appears that supersymmetry
cannot be achieved in general conformal field theory.
\par
Surprisingly, I was able to show \r\one\ that there is a very
general supersymmetry projection, which works in consistency
with modular invariance and can be implemented in any theory with
$N=2$ superconformal invariance, leading to space--time supersymmetry.
The supersymmetry charge is given
by $Q=\exp(i\phi)$, where $J=\partial_z\phi$ is the $U(1)$ current
algebra part of the $N=2$ superconformal algebra. In addition, one
demands that the total $U(1)$ charge, in both the compactified and
uncompactified dimensions, should be an integer. The crucial point
is that under the modular transformation $\tau\rarrow -{1\over\tau}$,
these two conditions are exchanged and thus can be implemented
simultaneously in any $N=2$ superconformal field theory, without
ruining modular invariance\footnote{\dagger}{The old GSO \r\GSO\ projection
is actually a particular case of the new G projection. The fact that
the GSO projection can be written in terms of an $N=2$ algebra
is known for some time, and was used in orbifold
calculations \REF\Orbifuck{L. Dixon, D. Friedan, E. Martinec, and
S.H. Shenker, Nucl. Phys. B282 (1987), 13; J.J. Atick, L. Dixon, and
A. Sen, Nucl. Phys. B292 (1987), 109}\r\Orbifuck . Preliminary
observations that the $G$ projection might be possible were reported
in \REF\Boucher{W. Boucher, D. Friedan and A. Kent, Phys. Lett.
B172 (1986), 316}\r\Boucher .}.
This new projection, the
$G$ projection, then leads to space--time supersymmetry in any
$N=2$ superconformal field theory.
\par
The next issue which needed to be addressed is how the left and
right movers are correlated in the string theory. Again, the constraint
of modular invariance is exceedingly restrictive. In general
conformal field theory, in order to be able to achieve modular invariance,
it is almost always required that the left movers and the right movers to be
identical `half theories'. In a heterotic--like \REF\het{D. Gross, J.
Harvey, E. Martinec and R. Rohm, Nucl. Phys. 265B (1985) 253}\r\het\
string theory, where
the left movers are fermionic and the
right movers are bosonic, this presents
a formidable problem, since the left and right movers, by definition,
are completely different.
\par
This problem is solved by a simple and completely general map
which takes any superstring theory into a heterotic--like string
theory \r\one . A general superstring theory in $d+2$ dimensions
contains a $d$ dimensional
flat superstring, described in the light--cone gauge by $d$ world
sheet free fermions and $d$ world--sheet free bosons. The world--sheet
fermions realize a level one $SO(d)$ current algebra, both in the
right and in the left moving sectors. The map that takes a superstring
into a heterotic string is then simply to replace the $SO(d)$
representations by $SO(24+d)$ or $E_8\times SO(8+d)$ ones in the right moving
sector, where one
exchanges the vector by the singlet and changes the sign of
the two spinors. The effect of this is to exchange the fermions that
carry a space--time index with fermions that carry an internal index.
This map preserves modular invariance
and spin--statistics and thus sends any consistent superstring theory in $d+2$
dimensions into
a consistent heterotic string in $d+2$ dimensions. In the case of
space--time supersymmetric
superstring compactification to four dimensions the resulting
gauge groups are either $SO(26)$ or $E_8\times E_6$. The $E_6$
is obtained by combining $SO(10)$ with the superconformal $U(1)$.
These gauge groups are the same as the ones obtained in the
supergravity models on CY manifolds \r\Candelas .
\par
In \r\two\ the following was demonstrated:

{\it String theory on a Calabi--Yau manifold exists, it
is a full fledged,
fully consistent string theory, which is space--time supersymmetric.
All string theories on a CY manifold have the structure described
above. In addition, any string theory which has this structure
is a string theory on some Calabi--Yau manifold.}

More generally, this structure corresponds to a string compactification
to $10-2k$ dimensions, with propagation on a manifold of $SU(k)$
holonomy for the cases $k=1,2,3$. Presently, the proof of the
above statement is incomplete. A partial proof and additional conclusive
evidence are presented in \REF\Future{D. Gepner, Talk given in
the superstring conferences in Boulder, August 1987, and in
Copenhagen, October, 1987,
to appear}\r\Future .

Our main tool in exploring this structure are the minimal $N=2$
superconformal field theories. The reason is that these are
the only presently known $N=2$ superconformal field theories.
\footnote{\ddagger}{Except of course for the trivial
realization in terms of free fermions. This realization corresponds
to the case of flat tori \REF\Schwarz{J.H. Schwarz, Phys.Rep. 89
(1982), 223}\r\Schwarz\
and their discrete quotients (orbifolds) \REF\Orb{L. Dixon, J. Harvey,
C. Vafa,
and E. Witten, Nucl. Phys. B261 (1985), 620, B274, (1986),
285}\r\Orb .} The minimal models have the trace anomaly
$$c={3k\over k+2}\qquad{\rm for\ }k=1,2,\ldots,\infty.\e$$
The primary fields in the minimal models are labeled by three
integers for the left movers, $l$,$q$, and $s$, which obeys
$0\leq l\leq k$, $q$ which is defined
modulo $2(k+2)$ and $s$ which labels the sector and is defined
modulo $4$. In addition, the right movers carry another set of such
quantum numbers. We denote such a primary field by
$\Phi_{l,q,s,\bar l,\bar q,\bar s}$. The dimension and charge
of this field are then
$$\Delta={l(l+2)-q^2\over 4(k+2)}+{s^2\over8}+{\rm integer},\qquad
Q=-{q\over k+2}+{s\over2}+2({\rm integer}).\e$$
\par
One can get the correct total trace anomaly by an arbitrary
tensoring of these models. The number of possibilities for consistent
string theories
is enormous, at least several millions in the case of $c=9$
which corresponds
to a four dimensional string theory. It appears that by these
possibilities alone one can get all Calabi--Yau manifolds
up to diffeomorphisms \r\Future.
\par
The rule for computing the spectrum is simple: anything that can
appear should appear. One starts from any modular invariant
$N=2$ theory, and implements the $G$ projection, along with
the map into heterotic--like string theory. The massless states
in a tensor product of $c<3$ minimal theories can then be described
as all the states obeying the following conditions,

(C1) The left and right states have a total $U(1)$ charge which is
odd integral.

(C2) The states are either in the Ramond sector of all the sub-theories
or all in the Neveu--Schwarz sector.
\REF\On{D. Gepner, Nucl. Phys. B287 (1987), 111}

(C3) In each of the discrete models, the $l$ and $\bar l$ quantum numbers
are correlated according to any of the $A^1_1$ invariants, which were
classified in \r\On . The left and right $q$ and $s$ quantum numbers
are equal. These conditions guarantee modular invariance.

(C4) In addition, we add to the spectrum states which can be obtained
by the action of $Q$, the supersymmetry charge.
This condition implements space--time supersymmetry. We also add
to the spectrum states related by $G_iG_j$ where the $G_i$ are the
superconformal stress tensors in any of the sub-theories
\par
In order to identify these string theories as string propagation
on Calabi--Yau manifolds we explored the massless spectrum.
In the supergravity models the gauge group is $E_8\times E_6$,
the number of generations ($27$ of $E_6$) is equal to $h^{2,1}$
and the number of anti--generation is $h^{1,1}$ \r\Candelas . If, in addition,
the manifold has some automorphism group (this is the physically interesting
case) then, by equivariant index theorems, the generations and
anti--generations
must transform in as the forms. By comparing the automorphisms with the
discrete symmetries of the string theory and the way the massless spectra
transform, we then obtain a highly unambiguous, model by model, identification
of string theories with the spectrum expected for particular manifolds.
An example of this procedure will be described later in the context
of the three generations case.
\par
As a result we find that the massless spectrum of a string theory on
a Calabi--Yau manifold is as following,

1) The gauge symmetry is $E_8\times E_6$ or $SO(26)$ times a possible
extra gauge symmetry.

2) The theories have $N=1$ space--time supersymmetry.

3) The number of generations is $h^{2,1}$ and the number of anti--generations
is $h^{1,1}$.

4) The $E_6$ singlets are divided according to: singlets that perturb the
complex structure (their number is $h^{2,1}$), singlets that change the
radii (their number is $h^{1,1}$), singlets coming from
$H^1({\rm End}\,T)$ and a number of Higgs singlets equal in number
to the dimension of the extra gauge symmetry.

5) The automorphisms of the surface appear as discrete symmetries of the
string theories and the massless spectra transform as
their corresponding forms.

The physically interesting case is when the number of
generations is three, arising when the manifold has the Euler
number $|\chi|=6$.
Models with more than three generations tend to have
problems
with fast proton decay, as well as the flow of
coupling constants \REF\Phen{E. Witten, Nucl. Phys. B258 (1985), 75;
M. Dine, V. Kaplunovsky, M. Mangano, C. Nappi and N. Seiberg, Nucl.
Phys. B259 (1985) 549}\r\Phen, and thus can probably be ruled out.
\par
The first examples of CY manifolds with $\chi=6$
\REF\TY{S.T. Yau, Proceedings of the
Argonne Symposium on anomalies, geometry, and topology, eds. W.A.
Bardeen
and A.R. White (World Scientific, Singapore, 1985).}
were described by Tian and Yau \r\TY .
In refs.
\REF\Search{P. Candelas, A.M. Dale,
C.A. Lutken, and R. Schimmrigk, Preprint, UTTG-10-87, CERN-TH 4694-87
(1987)}\REF\Asp{P.S. Aspinwall, B.R. Greene, K.H. Kirklin, and P.J.
Miron, Oxford preprint, February 1987}\r{\Search,\Asp},
a comprehensive computer search for all complete
intersection manifolds with $|\chi|=6$ was carried out,
and it was shown that no additional such
manifold exists.
It is also known that there are no orbifolds \r\Orb, which corresponds
to propagation
on a flat singular limit of some CY manifold (e.g. the
Z manifold \r\Candelas), that have three generations.
\REF\Sch{R.
Schimmrigk, preprint NSF-ITP-87-34, UTTG-15-87}
In addition, all the known manifolds \r{\TY,\Sch}\ with $\chi=-6$
are
\REF\GK{B.R. Greene and K.H. Kirklin, Harvard preprint HUTP-87/A025
(1987)}
either diffeomorphic to one another, or ill defined \r\GK.
\par
Thus, strikingly, there appears to be
a {\bf unique\ }manifold with three generations, the Tian--Yau manifold.
In this paper we describe a string theory on this
manifold.
\par
Our starting point is a CY manifold with Euler number $\chi=-54$.
It can be described as the hypersurface, $S$, in $CP^2\times CP^3$,
described as the manifold of solutions of the polynomial equations
$$\eqalign{P_1&=z_0^3+z_1^3+z_2^3+z_3^3=0,\cr
P_2&=z_1x_1^3+z_2x_2^3+z_3x_3^3=0,\cr}\e$$
where $[z_0,z_1,z_2,z_3]\in CP^3$ and $[x_1,x_2,x_3]\in CP^2$.
This manifold has a vanishing first Chern class as follows from
the existence of a holomorphic three form, which can be written as
$$\mu=\oint \oint{\epsilon_{ijkl}z_i\d z_j \wedge\d z_k
\wedge\d z_l\,\epsilon_{ijk}
x_i\d x_j\wedge\d x_k\over P_1P_2},\e$$
where the integrals are taken around close contours
surrounding the surfaces $P_1=0$ and $P_2=0$.
\par
In order to compute the Hodge numbers for this manifold it is enough to
find $h^{2,1}$, since $\chi=2(h^{1,1}-h^{2,1})$. Now,
the Hodge number $h^{2,1}$ counts the number of deformations of the
complex structure in manifolds which admit a metric of $SU(3)$
holonomy.
The reason is that these deformations are given, in general,
by $(1,0)$ forms
with values in the tangent bundle, $H^1(T)$ (e.g. see \REF\Kod{K. Kodaira,
Complex
manifolds and deformations of complex structure, Springer--Verlag (1985).}
\r\Kod), which can in turn be converted
to $(1,2)$ forms using the holomorphic three form. The deformations of
the
complex structure for the particular surface $S$ may all be described
as perturbations of the defining equation (3). We may perturb $P_1$ by adding
any of the $20$ polynomials which are cubic in $z$ and of zero order in $x$,
or perturb
$P_2$ by any of the $40$ polynomials which are linear in $z$ and cubic in $x$.
In total there are $60$ possible polynomials. However, polynomials related
by a linear redefinition of $z$ or $x$ correspond to the same complex
structure.
There are $25$ such redefinitions and thus the net number of perturbations
is $h^{2,1}=35$. Since $\chi=-54$ we also find $h^{1,1}=8$.
\par
The surface $S$ enjoys a large global automorphism group. First, we can
permute the indices $1,2,3$ by an arbitrary permutation, $p\in S_3$, of
these indices:
$z_i\rarrow z_{p(i)}$ simultaneously with $x_i\rarrow x_{p(i)}$. Next, we
have a $Z_3\times Z_9^3$ automorphism group given by different
phases.
Denoting
by $\{r_0,r_1,r_2,r_3\}$ an element of $Z_3\times Z_9^3$, where $r_0$ is
defined modulo $3$ and the other
$r$ modulo $9$, its action is given by
$$z_i\rarrow e^{2\pi ir_i/3} \qquad {\rm for\ }i=0,1,2,3 \e$$
$$x_i\rarrow e^{-2\pi i r_i/9} x_i \qquad {\rm for\ } i=1,2,3.\e$$
Since an overall phase is irrelevant in $CP^n$, the group
element $g=\{1,1,1,1\}$ acts trivially. To summarize, the global automorphism
group is $G=S_3\semidirect(Z_3\times Z_9^3)/(g)$. It is of order $1458$.
\par
Under the automorphism group $G$ the
deformations of the complex structure transform like their corresponding
polynomial perturbations. We denote by a column vector the perturbations,
where the up (down) component perturb $P_1$ ($P_2$). Due to the freedom
to linearly redefine $z$, the perturbations of $P_1$ may all be assumed
to be linear in any of the $z_i$. Similarly, redefinitions of $x$ allow
us to write the perturbations of $P_2$ as $z_i x_j^2 x_k$, $z_i x_j^3$,
or $z_i x_1x_2x_3$, where $i\neq j$. The possible
perturbations then come
in the patterns,
$$\eqalignno{\pmatrix{z_0z_1z_2\cr0\cr}&\qquad (1,3,3,0)&{(3)\qquad}\cr
\pmatrix{z_1z_2z_3\cr0\cr}&\qquad (0,3,3,3)&{(1)\qquad}\cr
\pmatrix{0\cr z_0x_1^3\cr}&\qquad (1,6,0,0)&{(3)\qquad}\cr
\pmatrix{0\cr z_0x_1^2x_2\cr}&\qquad (1,-2,-1,0)&{(6)\qquad}\cr
\pmatrix{0\cr z_0x_1x_2x_3\cr}&\qquad (1,-1,-1,-1)&{(1)\qquad}\cr
\pmatrix{0\cr z_1x_2^3\cr}&\qquad (0,3,6,0)&{(6)\qquad}\cr
\pmatrix{0\cr z_1x_2^2x_3\cr}&\qquad (0,3,-2,-1)&{(6)\qquad}\cr
\pmatrix{0\cr z_1 x_2^2 x_1\cr}&\qquad (0,2,-2,0)&{(6)\qquad}\cr
\pmatrix{0\cr z_1 x_1 x_2 x_3\cr}&\qquad (0,2,-1,-1)&{(3)\qquad}\cr}$$
where we denote by $(m_0,m_1,m_2,m_3)$  the charge of a vector, $v$, in
any of the one dimensional
irreducible representation of $Z_3\times Z_9^3$. A vector in this
representation
transforms as
$$v\rarrow e^{2\pi i(r_0m_0/3+r_1m_1/9+r_2m_2/9+r_3m_3/9)}
v\e$$
under $\{r_0,r_1,r_2,r_3\}\in G$.
Since $g=(1,1,1,1)\in G$ is equivalent to $0\in G$, it must act trivially
in all the representations of $G$ and thus
$$3m_0+m_1+m_2+m_3=0\mod 9,\e$$
for all the representations $(m_0,m_1,m_2,m_3)$.
\par
In addition, the holomorphic three form $\mu$ transforms in the
representation $(1,2,2,2)$ of $G$.
The subgroup of $G$ which commutes with supersymmetry, $H$,
is given by the elements that act trivially on the holomorphic three
form,
$$H=\big\{\{r_0,r_1,r_2,r_3\}\in G\,\big\vert\,\, 3r_0+2r_1+2r_2+2r_3=0
\mod 9 \big\}.\e$$
All the permutations commute with supersymmetry.
The other elements of $G$, which are not in $H$, are $R$ symmetries.
\par
The $(2,1)$ forms are obtained from the deformations of the complex
structure, which are elements of $H^1(T)$, by multiplying with
holomorphic three form. Thus, the $(2,1)$ forms transform like the
deformations times the holomorphic three form, i.e. they differ
by the charge $(1,2,2,2)$.
\par
The transformation properties of the $(1,1)$ forms under $G$ may be
computed using Lefshets fixed point theorem. Let $f$ be some element of
the automorphism group $G$. Then $f$ acts on the cohomology group as some
matrix, $f^*$. Lefshets fixed point theorem tells us that
$$\sum_{p,q} (-1)^{p+q}\Tr_{H^{(p,q)}} f^*=\chi(M_f),\e$$
where $\chi(M_f)$ is the Euler character of the submanifold which is fixed
by $f$, $M=\{x\in M\vert f(x)=x\}$. By calculating all the Euler
numbers in eq. (10), we find that
the eight $(1,1)$ forms transform as
$$(0,0,0,0) \times 2,\qquad (1,3,6,6),\qquad (2,3,3,6).\e$$
\par
Let us turn now to one forms with values in the endomorphism of the tangent
bundle, $H^1({\rm End\,} T)$. In the field theory, such forms give
rise to massless $E_6$ singlets \REF\Witten{E. Witten, Nucl.Phys. 268B
(1986) 79}\r\Witten . These forms
are in correspondence with deformations of the tangent bundle.
Denote a tangent vector by $(U_a,V_b)$, where $U_a$ is a tangent vector
of $CP^3$ (a=0,1,2,3), and $V_b$ is a tangent vector in $CP^2$,
$b=1,2,3$. The tangent vectors $U_a$ and $V_b$ are defined modulo the
equivalence relation
$$U_a\sim U_a+\lambda z_a,\qquad V_b\sim V_b+\rho x_b,\e$$
for any $\lambda$ and $\rho$.
In addition, the vector $(U_a,V_b)$ must be tangent to the
two surfaces $P_1$ and $P_2$. This implies,
\def\parr#1#2{{\partial #1\over\partial #2}}
$$\eqalign{&\parr{P_1}{z_a}U_a=\parr{P_1}{x_b}V_b=0\cr
           &\parr{P_2}{z_a}U_a=\parr{P_2}{x_b}V_b=0\cr}\e$$
\par
A simple method to deform the tangent bundle is to change equation
(13)
by adding to it some small perturbation. We can perturb any of the
partial derivatives in (13) by adding to it an arbitrary polynomial
of the same bi-degree as the corresponding partial derivative.
Denote a perturbation by the matrix
$$M=\pmatrix{P_a&Q_b\cr L_a&R_b}.\e$$
Eq (13) is then perturbed by $(\parr{P_1}{z_a}+P_a)U_a=0$, etc.
The bi-degrees of the polynomials $P$, $Q$, $L$ and $R$ are $(2,0)$,
$(0,0)$, $(0,3)$ and $(1,2)$, respectively. In addition, the equivalence
relation eq. (12) implies
$$P^az_a=Q^bx_b=L^az_a=R^bx_b=0.\e$$
Any such set of polynomials defines a perturbation of the tangent
bundle.
All the perturbations come either from $P$ or from $R$. To perturb $P$
we may take $P_a=C_a P/z_a$, for an arbitrary $P$ which is of bi-degree
$(3,0)$ and where the $C_a$ are some constants which obey $\sum C_a=0$.
The possible $P$ come in the patterns
$$\eqalignno{z_0^2z_1&\qquad (2,3,0,0)&     {(3)\qquad}\cr
           z_0z_1^2& \qquad (1,6,0,0)&     {(3)\qquad}\cr
           z_1^2z_2& \qquad (0,6,3,0)&     {(6)\qquad}\cr
           z_0z_1z_2& \qquad (1,3,3,0)&     {(2\times 3)\qquad}\cr
           z_1z_2z_3& \qquad (0,3,3,3)&     {(2\times 1)\qquad}\cr}$$
where the numbers above denote the $Z_3 Z_9^3$ charges and multiplicities.
\par
Similarly, the perturbations of $R$ can be written as, $R_b=C_b R/x_b$,
where the constants $C_b$ obey, $\sum C_b=0$, and $R$ is any polynomial
of bi-degree $(1,3)$. The possible $R$'s fall into the patterns
$$\eqalignno{z_0x_1^2x_2& \qquad (1,-2,-2,0) &     {(6)\qquad}\cr
           z_1x_1^2x_2& \qquad (0,1,-1,0) &     {(6)\qquad}\cr
           z_2x_1^2x_2& \qquad (0,-2,2,0) &     {(6)\qquad}\cr
           z_3x_1^2x_2& \qquad (0,-2,-1,3) &     {(6)\qquad}\cr
           z_0x_1x_2x_3& \qquad (1,-1,-1,-1) &     {(2\times 1)
\qquad}\cr
           z_1x_1x_2x_3& \qquad (0,2,-1,-1) &     {(2\times 3)\qquad
}\cr}$$
In total we find $52$ elements of $H^1({\rm End\,}T)$.
There can be more deformations which cannot be obtained in this way.
Using spectral sequences it should be possible to compute the entire
cohomology.

In the field theory limit the number of generations ($27$ of $E_6$) is
$35$, corresponding to the $35$ harmonic $(2,1)$ forms, and the number of
anti--generations ($\bar{27}$ of $E_6$) is $8$, corresponding to the
harmonic $(1,1)$ forms. The net number of generations is $\half\vert\chi
\vert=27$.
\par
Consider now the theory made by gluing one copy of the $k=1$
model with three copies of the $k=16$ model. In addition, in condition
(C3) we use the sporadic modular invariant at level $16$
\r\On .
The resulting spectrum may be easily computed from (C1--C4). The
theory, denoted by $1^116^3$, contains $35$ generations ($27$ of $E_6$),
$8$ anti--generations ($\bar{27}$ of $E_6$) and $197$ massless $E_6$
singlets.
\par
As will be seen the theory $1^116^3$ corresponds to a string theory
on the manifold $S$. The number of generations and anti--generations
are indeed the same as $h^{2,1}$ and $h^{1,1}$ for this manifold.
\par
What are the discrete symmetries of the theory $1^116^3$?
The $k$'th minimal model has a $Z_{k+2}$ discrete symmetry. Thus,
the theory $1^116^3$ has a $Z_3\times Z_{18}^3$ symmetry. In addition,
we can permute the three identical $k=16$ sub--theories. However,
by examining the massless
spectrum, it becomes clear that each of the $Z_2$ subgroups of
$Z_{18}$ acts trivially on the spectrum and thus may be ignored.
Hence the symmetry group of the theory $1^116^3$ is $G=(Z_3\times S_3
\semidirect Z_9^3)/Z_9$. Denote an element of $Z_3\times Z_9^3$ by
$\{r_0,r_1,r_2,r_3\}$. The quotient by $Z_9$ corresponds to the
fact that the total superconformal $U(1)$ charge of all the fields
is an odd integer, implying that the element $\{1,1,1,1\}$ acts trivially on
all the fields in the theory, so the actual symmetry
group is a quotient
by the $Z_9$ subgroup generated by this element.
\par
We see that the theory $1^116^3$ has a symmetry group which is
isomorphic to the automorphism group of the hypersurface $S$.
\par
Under the $Z_{k+2}$
charge of the $k$ minimal model a field in the theory, $\Phi_{l,q,s,\bar l,
\bar q,\bar s}$, has a charge which is
$$Q=(q+\bar q)/2 \mod (k+2).\e$$
We assume
in this definition that $q=\bar q\mod2$. This does not create any
problem for the non R symmetries.
For some of the $R$ symmetries, however, eq. (16) may imply that
the charges are ill defined, suggesting that these group elements
are bad symmetries
that should be ignored. In the case at hand, though,
no such problem arises.
The R symmetries, since they do not commute with space--time
supersymmetry, are very tricky. Different supersymmetry partners transform
differently under them, and similarly, the different representations of
$SO(10)$, which make the $27$ or $\bar{27}$ of $E_6$, transform differently.
\par
We would like to compare the transformation properties of the various
massless fields in the spectrum of the $1^116^3$ theory, with those that
are predicted in the field theory.
\par
The first thing we note is that the plus and minus chirality components
of the $E_6$ gluino field come from the $H^{0,0}$ and $H^{0,3}$
Dolbeault
cohomology groups or the positive chirality gluino corresponds to the
unique constant $(0,0)$ form and the negative chirality gluino corresponds
to the antiholomorphic $(0,3)$ form. The adjoint representation of
$E_6$ decomposes into SO(10) as $78=1+16+\bar{16}+45$. The different
$SO(10)$ representations always transform differently under the R symmetries.
Now, for a fixed $SO(10)$ representation, say the singlet, the different
$Z_{k+2}$ charges of the positive and negative chirality gluinos will
always differ by $1$. This is simply because these two modes are related
by the square of the supersymmetry charge , $Q^2$, which, in turn, carries the
$Z_{k+2}$ charge which is $1$ for all the sub--theories. On the other
hand, this ratio corresponds to the holomorphic $(3,0)$ form.
Thus, the holomorphic $(3,0)$ form always carry the $Z_{k+2}$
charge $1$,
when this charge is defined as in eq. (16).
\par
{}From eq. (4) we see that the $(3,0)$ form has the $Z_3\times Z_9^3$
charge which is $(1,2,2,2)$. On the other hand, the positive and
negative chirality
gluinos differ by a $Z_3\times Z_{18}^3$ charge which is $(1,1,1,1)$.
Using this correspondence we can `fix the normalization' of the
discrete charges.
The discrete charges of the automorphisms of the manifold $(m_0,m_1,m_2,m_3)$,
which are elements of $Z_3\times Z_9^3$, and the conformal field theory
charges $(Q_0,Q_1,Q_2,Q_3)$, which are defined according to (16) and are
elements of $Z_3\times Z_{18}^3$, are then seen to be related as
$$m_0=Q_0\mod 3,\qquad m_i=2Q_i\mod 9,\quad {\rm for\ }i=1,2,3.\e$$
\par
Next, we can check whether the generations and anti--generations transform
as they are supposed to, in the field theory limit.
Under the non R symmetries (i.e. the ones which commute with SUSY) the
generations and anti--generations must transform like their corresponding
$(2,1)$ and $(1,1)$ forms.
The R symmetries are trickier since, as discussed earlier, different
supersymmetry components transform under them differently. Which
component, then, should we compare with the forms? The answer is that
the correct component for supersymmetry multiplets in the $27$ or
the $\bar{27}$ of $E_6$ is the scalar (helicity zero)
which is a vector of $SO(10)$. The reason is that such scalars are related
to the $E_6$ singlets which perturb the radius (in the $\bar{27}$ case) or
deform the complex structure (in the $27$ case) \r\Future\
and thus must transform in precisely the same way as the forms
of $H^1(T)$ (for $27$) or the $(1,1)$ forms (for $\bar{27}$) do.
\par
The following is an enumeration of the $35$ generations in the
$1^116^3$ theory,
\def\p#1{\Phi_{#1}}\def\q#1{\Theta_{#1}}

%  Generations --- table 1
% -------------
$$\eqalignno{
% 201  (4)
% ( 1, 3,2) ( 4,32,0) (16,20,0) (16,20,0)
(3)&\quad\p{1,2,1,1,3,2}\q{12,13,1,4,32,0}\q{0,1,1,16,20,0}^2
& {     ( 1, 6,0,0) }\quad\cr
% ( 1, 2,1) (12,13,1) ( 0, 1,1) ( 0, 1,1)
%
% 288  (3)
% ( 1, 3,2) ( 8,28,0) (12,24,0) (16,20,0)
(6)&\quad\p{1,2,1,1,3,2}\q{8,9,1,8,28,0}\q{4,5,1,12,24,0}\q{0,1,1,16,20,0}
& {    (1,-2,-1,0) }\quad\cr
% ( 1, 2,1) ( 8, 9,1) ( 4, 5,1) ( 0, 1,1)
%
% 302   (2)
%( 1, 3,2) (10,26,0) (10,26,0) (16,20,0)
(3)&\quad\p{1,2,1,1,3,2}\q{6,7,1,10,26,0}^2\q{0,1,1,16,20,0}
& {    (1,3,3,0) }\quad\cr
%( 1, 2,1) ( 6, 7,1) ( 6, 7,1) ( 0, 1,1)
%
% 314   (1)
%( 1,3,2) (12,24,0) (12,24,0) (12,24,0)
(1)&\quad\p{1,2,1,1,3,2}\q{4,5,1,12,24,0}^3
& {    (1,-1,-1,-1)}\quad\cr
%( 1, 2,1) ( 4, 5,1) ( 4, 5,1) ( 4, 5,1)
%
% 431   (9)
%( 0, 2,2) ( 4,32,0) (10,26,0) (16,20,0)
(6)&\quad\p{0,1,1,0,2,2}\q{12,13,1,4,32,0}\q{6,7,1,10,26,0}\q{0,1,1,16,20,0}
& {    (0,6,3,0) }\quad\cr
%( 0, 1,1) (12,13,1) ( 6, 7,1) ( 0, 1,1) (
%
% 458   (8)
%( 0,2,2) ( 6,30,0) ( 8,28,0) (16,20,0)
(6)&\quad\p{0,1,1,0,2,2}\q{10,11,1,6,30,0}\q{8,9,1,8,28,0}\q{0,1,1,16,20,0}
& {    (0,2,-2,0) }\quad\cr
%( 0, 1,1) (10,11,1) ( 8, 9,1) ( 0, 1,1) (
%
% 462  (7)
%( 0,2,2) ( 6,30,0) (12,24,0) (12,24,0)
(3)&\quad\p{0,1,1,0,2,2}\q{10,11,1,6,30,0}\q{4,5,1,12,24,0}^2
& {   (0,2,-1,-1) }\quad\cr
%( 0, 1,1) (10,11,1) ( 4, 5,1) ( 4, 5,1) (
%
% 472   (6)
%( 0,2,2) ( 8,28,0) (10,26,0) (12,24,0)
(6)&\quad\p{0,1,1,0,2,2}\q{8,9,1,8,28,0}\q{6,7,1,10,26,0}\q{4,5,1,12,24,0}
& {    (0,-2,3,-1) }\quad\cr
%( 0, 1,1) ( 8, 9,1) ( 6, 7,1) ( 4, 5,1)
%
%478   (5)
%(0,2,2) (10,26,0) (10,26,0) (10,26,0)
(1)&\quad\p{0,1,1,0,2,2}\q{6,7,1,10,26,0}^3
& {    (0,3,3,3) }\quad\cr
%( 0, 1,1) ( 6, 7,1) ( 6, 7,1) ( 6, 7,1)
%
}$$
% end table 1
The fields in the list correspond to anti--spinors which are singlets
of $SO(10)$. We denoted by $\Phi$ and $\Theta$ the fields from the $k=1$
and $k=16$ theories. The six indices on each field correspond to the
three left quantum numbers $(l,q,s)$ and the three right quantum numbers.
The numbers on the right are the $Z_3 Z_9^3$ charges of each of the fields.
The $Z_9$ charges are computed according to $m_i=q_i+\bar q_i-3\mod9$, for
$i=1,2,3$; the $Z_3$ charge is $m_0=-q_0-\bar q_0\mod3$.
We see that the $35$ generations transform in precisely the same representation
of
$G$ as the deformations of the complex structure do (p. 10).
\par
We can now check the anti--generations. The $8$ anti--generations,
along with their corresponding $Z_3\times Z_9^3$ charges, are enumerated
below,
%  Anti-generations: --- table 2
% ------------------
$$\eqalignno{
%
%*** 295
%( 0, 0,0) ( 8, 8,0) (14,14,0) (14,14,0)
(3)\quad&\p{0,1,1,0,0,0}\q{2,3,1,8,8,0}\p{8,9,1,14,14,0}^2
& {   (1,3,6,6)}\quad\cr
%( 0, 1,1) ( 2, 3,1) ( 8, 9,1) ( 8, 9,1)
%
%**** 315
%( 1,3,2) (12,12,0) (12,12,0) (12,12,0)
(1)\quad&\p{1,2,1,1,3,2}\q{4,5,1,12,12,0}^3
& {    (0,0,0,0)}\quad\cr
%( 1, 2,1) ( 4, 5,1) ( 4, 5,1) ( 4, 5,1)
%
%*** 615
%( 1, 1,0) ( 8, 8,0) ( 8, 8,0) (14,14,0)
(3)\quad&\p{1,2,1,1,1,0}\q{2,3,1,8,8,0}\q{2,3,1,8,8,0}\q{8,9,1,14,14,0}
& {    (2,3,3,6)}\quad\cr
%( 1, 2,1) ( 2, 3,1) ( 2, 3,1) ( 8, 9,1)
%
%*** 622
%( 0,4,2) (10,10,0) (10,10,0) (10,10,0)
(1)\quad&\p{0,1,1,0,4,2}\q{6,7,1,10,10,0}^3
& {    (0,0,0,0)}\quad\cr
%( 0, 1,1) ( 6, 7,1) ( 6, 7,1) ( 6, 7,1)
%
}$$
The fields above are also anti--spinors which are $SO(10)$ singlets.
The $Z_3\times Z_9$ charge, $(m_0,m_1,m_2,m_3)$ is computed in this
case by $m_0=-q_0-\bar q_0-1\mod3$, $m_i=q_i+\bar q_i+1\mod9$.
Again, we see that the anti--generations transform in precisely
the same way as the $(1,1)$ forms do, eq. (11). This completes the
identification
of the $1^116^3$ theory as a string theory on the hypersurface $S$.
\par
We can further compare the $E_6$ singlets. The $197$ singlets can be
seen to contain, in a completely unambiguous way, $35$ modes which
transform like $H^1(T)$, these are the singlets related to deformations
of the complex structure, $8$ modes transforming like $(1,1)$ forms,
these are singlets related to change of radii and $52$ singlets
transforming like the modes of $H^1({\rm End}\, T)$ described
earlier (p. 13).
The remaining
$102$ singlets may correspond to additional perturbations of the tangent
bundle that we have not computed, or less likely, to accidental massless
particles. The resolution of this question must await the complete
calculation of $H^1({\rm End}\,T)$ for this manifold.
\par
The automorphism group of $S$ has a certain $Z_3\times Z_3$ subgroup, $H$,
which will be very important for us. The first $Z_3$ is generated by
the permutation: $z_1\rarrow z_2\rarrow z_3\rarrow z_1$ along
with $x_1\rarrow x_2\rarrow x_3\rarrow x_1$, denoted by $h$.
This can be seen to be
a freely acting automorphism. The second $Z_3$ is generated by the
$Z_3\times Z_9^3$ group element $g=\{0,3,6,0\}$. This $Z_3$ is not
freely acting. Thus, the quotient manifold $S/H$ is a singular manifold.
However, these singularities can be resolved,
as discussed in ref. \r\Sch\
and the resulting
manifold is a CY manifold with
Euler number $\chi=-6$. Thus, a string theory on $S/H$ should have
three generations.
\par
The spectrum of a heterotic string theory propagating on the manifold
$S/H$ can be computed as a quotient of the theory $1^116^3$. The partition
function of the $k$'th minimal model, twisted in the space and time
directions by the $Z_{k+2}$ elements $x$ and $y$, respectively, is given
by \r{\one,\two}
$$Z(x,y)={1\over2}e^{2\pi ixy/(k+2)}\sum_{l,q,s} e^{2\pi ixq/(k+2)}
\chi^l_{q+2y}\chi^{l*}_{q,s},\e$$
where $\chi_{q,s}^l$ is the partition function of the $N=2$ conformal
block with the quantum numbers $(l,q,s)$.
Implementing eq. (18) in the string theory amounts to a simple
modification of the conditions (C1--C4) and enables us to compute
the spectrum of string propagation on the manifold $S/H$.
\par
Consider first string theory on the quotient manifold $S/(g)$.
By a straightforward enumeration of states we find that this theory
has $23$ generations, $14$ anti--generations and $173$ singlets.
Of these, $17$ generations, $8$ anti--generations and $85$ singlets
come from the untwisted sector.
\par
The Hodge numbers $h^{2,1}=23$
and $h^{1,1}=14$ are the same as those of a well known CY
manifold, namely
the one constructed by Tian and Yau
\r\TY . This manifold
can be described as the intersection of three hypersurfaces of
bi-degrees $(3,0)$, $(0,3)$ and $(1,1)$ in the product space
$CP^3\times CP^3$. Its most symmetric shape is
$$\sum_{i=0}^3 x_i^3=0,\qquad\sum_{i=0}^3 y_i^3=0,\qquad
           \sum_{i=0}^3 x_i y_i=0.\e$$
This manifold has the Hodge numbers $h^{2,1}=23$ and $h^{1,1}=14$.
Thus, taking a quotient of it by the freely acting $Z_3$ automorphism
group, which is generated by
$$(x_0,x_1,x_2,x_3)\times (y_0,y_1,y_2,y_3)
\rarrow (x_0,\alpha^2x_1,\alpha x_2,\alpha x_3)\times (y_0,\alpha y_1,
\alpha^2y_2,\alpha^2 y_3),\e$$
where $\alpha=\exp(2\pi i/3)$,
we get to a three generation manifold. The manifold $S/H$ is indeed
diffeomorphic to the Tian--Yau manifold \r\GK.
\par
The supergravity model on the Tian--Yau manifold was studied by a
number of authors \REF\Ross{B.R. Greene, K.H. Kirklin,
P.J. Miron, and G.G. Ross, Nucl.Phys. B278 (1986) 667; and preprint,
February 1987; S. Kalara and R.N. Mohapatra, Univ. of Minnesota preprint
UMN-TH-590-86; P. Candelas and
S. Kalara, preprint UMN-TH-601-87, UT-TH-12/87}\r\Ross\ and the
indications are that the discrete symmetries of the manifold
may well be rich enough to prevent fast proton decay.
\par
Returning to the string theory on $S$, the next step after twisting the
theory by $g$, is to further twist it by the permutation $h$.
By projecting
the spectrum of the string theory on $Q/(g)$ onto the $h$ invariant subspace,
it is easy to write the spectrum of the closed string (untwisted)
sector.
We find in this sector $9$ generations, $6$ anti--generations
and $62$ singlets. In addition, one needs to take into account the winding
sectors.
Since $h$ acts freely, these sectors do not
contribute any generations or anti--generations.
Thus, all together, in this
string
theory we find $9$ generations and $6$ anti--generations, or a net
number
of three generations.
\par
The Yukawa couplings in this three generation string theory can
all be computed exactly since they are
given as products of
the structure constants of the $N=2$ minimal models. These, in turn,
are related to the structure constants of the $SU(2)$ WZW models
which have been studied
by several authors \REF\Yukawa{
A.B. Zamolodchikov and V.A. Fateev,
Sov.J.Nucl.Phys. 43 (1986) 657;
P. Christe and R. Flume,
Nucl. Phys. B282 (1987) 466;
J. Fuchs and D. Gepner, Nucl. Phys. B294 (1987) 30;
J. Fuchs, preprint HD-THEP-97-11,
to appear in Nucl.Phys.B}\r\Yukawa . Using the isomorphism of states
and vertex operators that create them out of the vacuum \r\Group,
one can express the vertex operators in this string theory in terms
of WZW fields and free bosons, and thus to calculate the Yukawa
couplings exactly (see \r{\one,\two}\ for more explanation).
\par
In conclusion, we have presented a new string theory which corresponds
to string propagation on a three generation Calabi--Yau manifold.
It is a full fledged string theory in which all
the physical predictions can be computed
exactly. This string theory appears to be the unique viable candidate in
its class.
In addition, it comes intriguingly close to a
realistic description of nature, which is moreover a consistent
unification of gravity.
\par
\ack
I thank G. Faltings and D. Gross for interesting discussions.
\refout
\bye